\documentclass[preprint,12pt]{elsarticle}

\usepackage{graphicx}
\usepackage{amssymb}

\usepackage{lineno}

\journal{Journal of Communications Technology and Electronics}

\begin{document}

\begin{frontmatter}


\title{Reflection of Electrons during Tunneling and an Intersubband Polaron in the 2D Electron System
of a Delta-Layer in GaAs}



\author{S. E. Dizhur}
\address{Institute of Radioengineering and Electronics RAS, Moscow, Russia}

\author{I. N. Kotel'nikov}
\address{Institute of Radioengineering and Electronics RAS, Moscow, Russia}

\author{E. M. Dizhur}
\address{Institute for High Pressure Physics RAS, Troitsk, Moscow, Russia}

\begin{abstract}
Al-$\delta$-GaAs structures are studied where tunneling to one or more subbands of the 2D electron system
of a near-surface delta-doped layer is observed. 
Reflection of electrons at the threshold of the emission of an LO-phonon is observed. 
This phenomenon occurs when a new subband is involved in the tunneling process and when intersubband transitions of the electrons in a 2D-system with the emission of an LO-phonon are added
to the inelastic tunneling within a single subband. 
It is shown that, under the conditions of an intersubband polaron resonance, the reflection processes dominate during tunneling into the delta-layer. 
If, however, tunneling from the delta-layer occurs, the reflection processes are observed when two subbands are occupied.
\end{abstract}


\end{frontmatter}


\section{INTRODUCTION}
\label{S:intro}

The technology of delta-doping (hereinafter, $\delta$-doping) enables implementation of extremely inhomogeneous doping in GaAs by the method of molecular beam epitaxy (MBE). 
Ideally, in this case, atoms of the base lattice are only replaced with doping atoms within
a single lattice plane, where $\delta$-doping is performed (see, e.g., review~\cite{1}). 
As a result, a V-shaped well for electrons is created in this plane (for example, in the
case of Si doping). 
The electrons are confined by the charges of donors near the $\delta$-doped layer. 
The conductivity of the 2D electron system (2DES) in a $\delta$-layer may be sufficiently high owing to both the high maximum concentration of carriers ($\simeq 10^{13}$\, cm$^{-2}$) and to the growth of their mobility that is observed as the 2D subband’s number increases. Such properties of the 2DES in
$\delta$-layers show that they potentially may be used as a field-effect transistor's channel~\cite{1,2,3}. If the distance between the $\delta$-doped layer and the metal/GaAs interface is reduced to tens of nanometers, the onset of the carriers’ transitions between the metal and the $\delta$-layer due to the tunnel effect is observed. 
Such a structure with a high-quality Al/GaAs interface produced in an MBE chamber opens a new avenue for the use of the near-surface $\delta$-layer. 
An Al-$\delta$-GaAs tunnel junction produced on the basis of this layer makes it possible to
analyze effects of the density of states and many-particle interactions in a 2DES by the tunnel spectroscopy
method~\cite{4}. 
Such a tunnel system seems the most promising for observation of the resonance phenomena related to the polaron interaction between different levels~\cite{5}. 
It is important that the distances between the quantum confinement levels in the $\delta$-doped layer in
GaAs prove to be close to the energy of longitudinal optical phonons, $\varepsilon_{LO} = 36.5$\, meV. 
The positions of levels in the 2D channel of the Al-$\delta$-GaAs junction may be altered using, e.g., a diamagnetic shift to bring the energy between subbands to resonance with $\varepsilon_{LO}$.
Such an experiment was carried out in~\cite{4}, where anticrossing of terms characteristic to interacting levels was observed in the case when $E_{01} = 2\varepsilon_{LO}$. 
Similar effects (anticrossing of terms) were observed in optical experiments involving tunnel-coupled quantum wells~\cite{6, 7} and in magnetic tunneling in AlGaAs/GaAs structures~\cite{8}. 
These effects are a manifestation of the interaction between the Landau levels of different subbands
when the ratio of the energy between the subbands and the cyclotron energy is an integer.

Al-$\delta$-GaAs structures exhibit one more important feature. 
In addition to the low-density 2DES, a 2DES with a very high density of electrons can be created, where Fermi energy $\varepsilon_{F\delta}$  is on the order of the LO-phonons’ energy or even exceeds it. 
Similar to tunneling between 3D electrodes, the 2DES spectrum in the latter case should exhibit singularities in the self-energy (related to polarons) at distances $\pm \varepsilon_{LO}$
from the Fermi surface~\cite{9}. 
Effects of inelastic tunneling during interaction with an LO-phonon~\cite{10} may be observed as
well and should not depend on the ratio between $\varepsilon_{F\delta}$ and $\varepsilon_{LO}$. 
The lines of LO-phonons have been, indeed, observed in Al-$\delta$-GaAs structures at the potential bias
$U = \pm \varepsilon_{LO}/e$, where $e$ is the electron charge. 
However, their form strongly depends on the variation of the intersubband energies near the polaron resonance~\cite{11} or on the variations in the spectrum of levels in a 2DES.
Additionally, it has been found that, when two subbands are involved in the tunneling process, a new type
of phonon lines is observed~\cite{12, 13}, which corresponds to a decrease in tunnel conductance (reflection
of electrons). 
When the concentration is increased in the 2DES and, hence, the number of the subbands
involved in tunneling increases, the process turns out to be more complicated: reflection of electrons at the threshold of the emission of LO-phonons may prove to be the dominant effect in the conditions of the polaron resonance between levels $E_{1}$ and $E_{2}$ ($E_{12} = \varepsilon_{LO}$) when
tunneling into a 2DES occurs. 
Under certain conditions, this reflection may be observed during tunneling from a 2DES.

\section{SAMPLES AND THE MEASUREMENT PROCEDURE}
\label{S:smpl_meas}

\begin{figure}[ht]
\centering\includegraphics[width=\columnwidth]{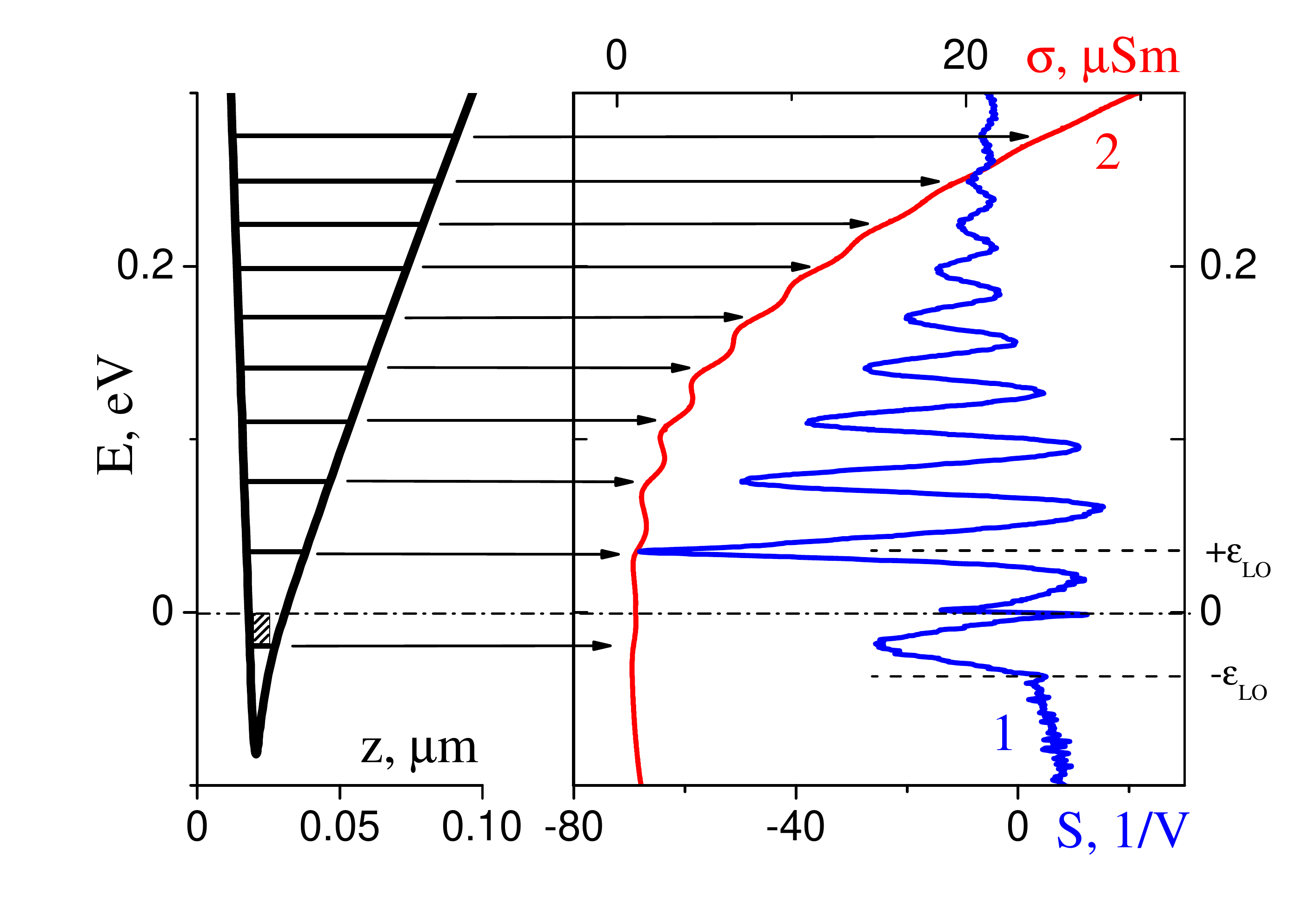}
\caption{
(a) Potential profile of a conductivity band's bottom in a semiconductor electrode of an Al-$\delta$-GaAs tunnel junction. 
Horizontal lines show positions of the ten first energy levels of $E_{i}$ in the $\delta$-doped layer, of which only one (the lowest) level, $_E0$, is occupied (crosshatched area). 
(b) Results of measurements of tunnel conductance $\sigma(E)$ (curve 1) and its logarithmic derivative 
$S = d\ln\sigma/dU(E)$ (tunnel spectrum, curve 2) for an Al-$\delta$-GaAs tunnel junction. 
(The area of the tunnel gate is $10^{-2}$\, mm$^2$.)
}
\label{1}
\end{figure}

The Al-$\delta$-GaAs structures produced at the Institute of Radio Engineering and Electronics of the Russian Academy of Sciences were studied. 
The samples were made by the molecular beam epitaxy described in~\cite{4}.
Owing to the technology where aluminum is deposited directly in the MBE chamber, the interface between Al
and GaAs had the best possible quality. 
Distance $z$ between this interface and the $\delta$-doped (Si) layer was $\sim ~20$\, nm. 
The tunnel spectra of the Al-$\delta$-GaAs junction were measured at a temperature of $4.2$\, K. 
The measurement procedure is described in~\cite{4}. 
The dependence of the logarithmic derivative $S = d \ln \sigma / dU$ of tunnel conductance
$\sigma = dI/dU$ on $U$ was used as a tunnel spectrum.
Such a choice made it possible to reduce the tunnel characteristics of junctions with different values of
$\sigma$ to a unified scale. 
The tunnel conductance in the case of tunneling from a 3D- to a 2D-state is known~\cite{14} to
be proportional to the density of states $\rho_{i}(E)$ in the 2DES:
\begin{equation}
\sigma(U)\simeq \sum_{i} D_{i}(U)\rho_{i}(U),    
\end{equation}
where $D_{i}(U)$ is the tunnel barrier transparency for level $E_{i}$. 
This is clearly seen in Fig.~\ref{1}, where potential profile $E(z)$ for an Al-$\delta$-GaAs tunnel junction is shown. 
This figure also displays measured dependence $\sigma(E_{Fm} - E_{F\delta})$, which exhibits step-like changes in the density of 2DES states when the Fermi level in metal $E_{Fm}$ crosses the
energy levels in the $\delta$-layer. 
This behavior is more pronounced in the tunnel spectrum $S(E) = d\ln\sigma/dU$, where
positions of levels $E_{i}$ correspond to smooth minima on curve $S(E)$. 
The arrows indicate step-like changes in the 2D density of states $\rho(E)$ observed in the tunnel conductance displayed as a function of the difference $E = E_{Fm} - E_{F\delta}$ between the Fermi levels of the electrodes.
They also show the dips in the tunnel spectrum whose location in the energy spectrum corresponds to the
position of the energy levels in the $\delta$-layer. 
In addition, the tunnel spectrum exhibits features that are typical of many-particle effects: phonon lines located at $E = \pm \varepsilon_{LO}$ and a zero-bias anomaly. 
The curves shown in Fig.~\ref{1} show the high quality of the Al-$\delta$-GaAs tunnel junction,
whose spectrum exhibits one occupied ($E_{0} < 0$) and nine empty quantum confinement subbands that
can be observed with high reliability. 
The energy $E_{iF} = E_{i} - E_{F\delta}$ can be observed using the tunnel spectra~\cite{4},
and the value of $|E_{0F}| = 20$\, meV corresponds to the Fermi energy of the 2DES.

\begin{figure}[ht]
\centering\includegraphics[width=0.6\columnwidth]{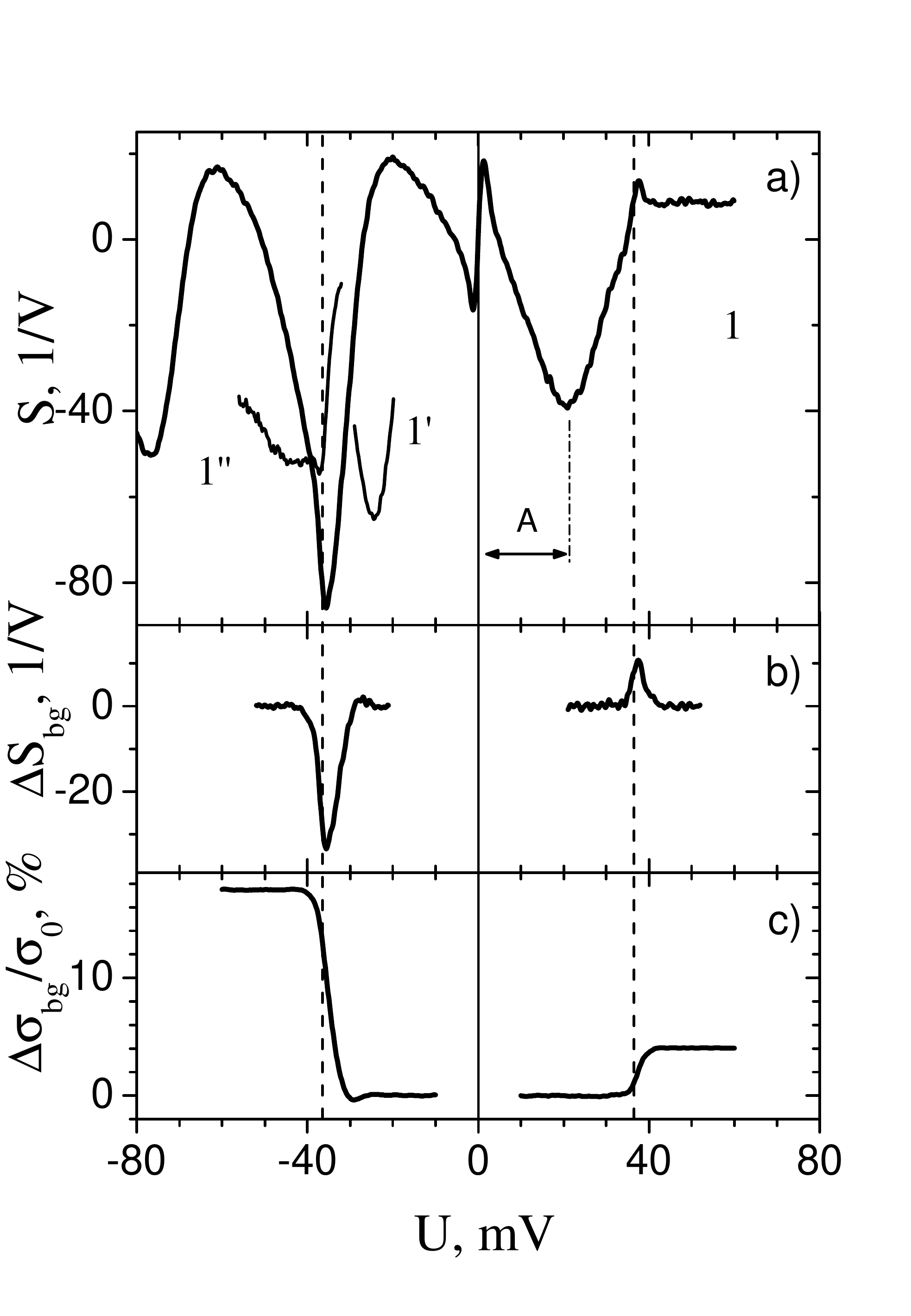}
\caption{
(a) Tunnel spectrum $S$ for sample 1, (b) quantity $\Delta S_{bg}$ obtained by subtracting background line $S_0$, and (c) relative conductivities $\Delta\sigma_{bg}/\sigma_0$ in the region of phonon singularities.
The figure also shows the parts of the spectrum near the phonon singularity at $U = -36.5$\, mV corresponding to (1') the shift of level $E_1$ after exposure to light from an LED over $12.8$\, s and (1") the diamagnetic shift of the same level in a magnetic field of $7.4$\, T directed along the $\delta$-layer. 
A shows the energy of the occupied level, $E_{0F}$, measured from the Fermi level.
}
\label{2}
\end{figure}

Figure~\ref{2} shows the tunnel spectrum of sample 1 ($4.2$\, K), which is cut from the same substrate material as the sample shown in Fig.\~ref{1}; however, the area of its tunnel gate is $0.5$,\ mm$^2$. Voltage $U$ across the junction, which is considered positive when electrons tunnel into
metal ($E_{Fm} < E_{F\delta}$), is equal to $eU = E_{F\delta} - E_{Fm}$. 
In the sample 1, when tunneling in the 2DES occurs at the threshold of the emission of an LO-phonon, i.e., at $U = -36.5$\, mV, electrons can transit to one or two subbands under the effect of the external influence described below. 
The Fermi level in metal, $E_{Fm}$, at the threshold of the emission of an LO-phonon proves to be higher
than the Fermi level in the $\delta$-layer ($E_{F\delta}$) by the value of $\varepsilon_{LO}$. 
Energies $E_{iF}$ of three lower subbands can be determined using the tunnel spectrum shown in Fig.~\ref{2}: $E_{0F} = -20$, $E_{1F} = 36$, and $E_{2F} = 76$\, meV. 
Figure\ref{2} also shows the lines of LO-phonons separated from the background.

\begin{figure}[ht]
\centering\includegraphics[width=0.8\columnwidth]{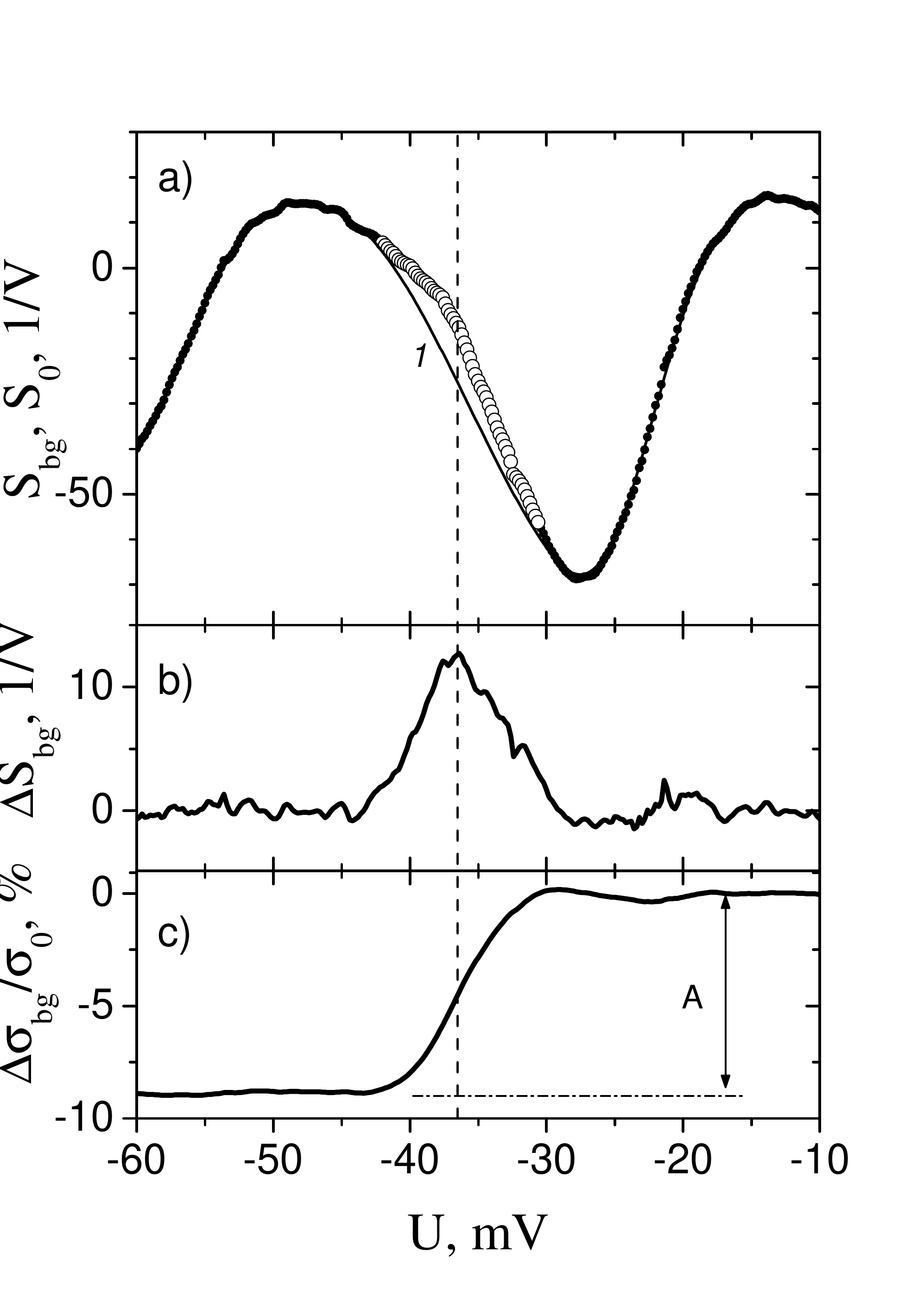}
\caption{
a), curve 1, Determination of background line $S_0$ in the part of background spectrum $S_{bg}$ (closed and open circles) and b) the result of separating the phonon line $\Delta S_{bg} = S_{bg} - S_0$. 
Open circles show a band of $\pm 6$\, mV around the center of the phonon line. 
Integration of $\Delta S_{bg}(U)$ over voltage $U$ yields c) the dependence of $\Delta\sigma_{bg}/\sigma_0$ on $U$, which is used to determine (A) the magnitude of step $\Delta\sigma^{*}$ of the tunnel conductance at the threshold of the emission of an LO-phonon.
}
\label{3}
\end{figure}

To determine background curve $S_0$ and to separate the lines of optical phonons from the tunnel spectra,
spectrum $S_{bg}$ outside the phonon-related singularity has been approximated by cubic splines (Fig.~\ref{3}). 
This procedure, which was successfully applied for the first time in~\cite{15}, is described in detail in~\cite{16}. 
The part of curve $S_{bg}(U)$ containing the singularity (range of the bias values $\pm 6$\, mV around the line center for $U = -36.5$\,mV) is taken into account by a weight that is significantly
smaller than that of the other points in the approximation region. 
Background line $\Delta S_{bg}$ is separated as the difference $\Delta S_{bg} = S_{bg} - S_0 = d(\ln \sigma_{bg} - \ln \sigma_0)/dU$. 
Assuming that the tunnel conductance represented by the formula $\sigma_{bg} = \Delta\sigma_{bg} + \sigma_0$, where $\Delta\sigma_{bg} \ll \sigma_{0}$, is the variation of the conductance due to the electron–phonon interaction, we arrive at $\Delta S_{bg} \simeq d(\Delta\sigma_{bg}/\sigma_{0})/dU$. Therefore, integration of $\Delta S_{bg}$ over $dU$ yields the relative variation of tunnel conductance
$\Delta\sigma_{bg}/\sigma_0$ in the vicinity of the phonon line (Fig.~\ref{3}c).
When separating the phonon line, the position of the line's center is chosen at $|U| = 36.5$\, mV. Therefore, the separated line $\Delta\sigma_{bg}/\sigma_{0}$ depends weakly on the method
used to determine $\sigma_0$ under reasonable assumptions of the width of the area where the phonon singularity is
significant.

The phonon line for sample 1 (Fig.~\ref{3}) was measured after having exposed the sample to light from an LED for $3.2$\, s. 
This procedure corresponds to the unusual processes of interaction between electrons and phonons
during of tunneling~\cite{12} because the conductance decreases when the threshold $|eU| = \varepsilon_{LO}$ is attained. 
This phenomenon is associated with a negative step in the dependence of $\Delta\sigma_{bg}/\sigma_0$ on $U$ (the magnitude of step $\Delta\sigma^{*} < 0$ in Fig.~\ref{3}). 
Inelastic processes involving emission of an LO-phonon are known (see, e.g.,~\cite{14}) to
increase conductivity in the region above the threshold, so that $\Delta\sigma^{*} > 0$. 
Lines of such type were observed in sample 1 after it had been cooled to $4.2$\, K in darkness
(see Fig.~\ref{2}). 
Below, we consider the dependence of the height of step $\Delta\sigma^{*}$ in the tunnel conductance on the
position of the bottom of the subband, which is closest to the energy $E_{F\delta} + \varepsilon_{LO}$. 
It will be shown that inversion of the sign of $\Delta\sigma^{*}$ can be observed in a sample by
changing the positions of 2D subbands under the effect of external influence.

\section{RESULTS AND DISCUSSION}
\label{S:res}

\subsection{Reflection of Electrons during Tunneling into a Low-Density 2DES}

\begin{figure}[ht]
\centering\includegraphics[width=0.6\columnwidth]{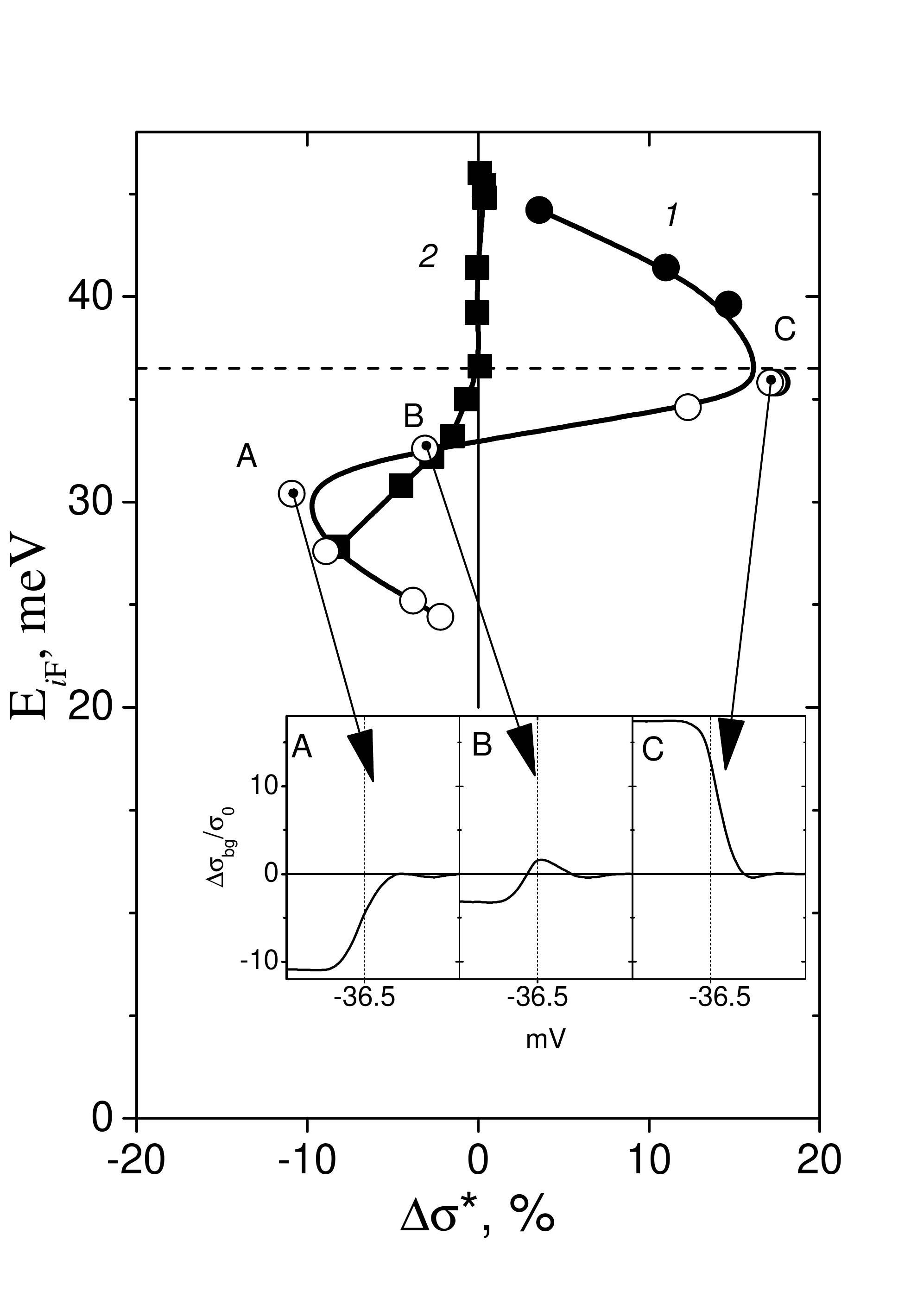}
\caption{
Experimental dependence of the magnitude of step $\Delta\sigma^{*}$ in tunnel conductance on the position of level $E_{1F}$ for sample 1 with a single occupied band. 
Filled circles correspond to the diamagnetic shift of the level and open circles
correspond to the level shift after lighting (PTPC mode);
$\Delta\sigma^{*} < 0$ corresponds to the decrease of tunnel conductance after the threshold. 
The horizontal dashed line indicates the energy of an LO-phonon. 
The inserts (A--C) show dependences $\Delta\sigma_{bg}/\sigma_0$ on $U$ at the respective points. 
Filled squares show the dependence of $\Delta\sigma^{*}$ on $E_{2F}$ for sample 2 with two
occupied subbands under the conditions of the diamagnetic shift of level $E_{2F}$.}
\label{4}
\end{figure}

In sample 1, the Fermi energy in a single occupied band is only $20$\, meV, a value that corresponds to a low-density 2DES (a concentration of 2D electrons of $\approx 5 \times 10^{11}$\, cm$^{-2}$). 
The case of the low density of 2D electrons in a $\delta$-layer is the most convenient for identifying the
mechanism of reflection of electrons ($\Delta\sigma^{*} < 0$) during tunneling to a 2DES since, according to Fig.~\ref{2}, no more than two subbands are involved in the tunnel current at the threshold of the emission of an LO-phonon. 
Let us use Fig.~\ref{4} to analyze the behavior of $\Delta\sigma^{*}$ in sample 1 as a function of the position of subband's bottom $E_{1F}$. 
Initially, the subband's bottom is located at $E_{1F} = 36$\, meV, i.e., near the threshold of the phonon emission, corresponding to point B. 
For $U = -36.5$\, mV, the tunnel current in this case mainly depends on the transition to
unoccupied states in subband $E_0$. 
Under these conditions, $\Delta\sigma^{*} > 0$ corresponding to usual processes of inelastic tunneling: opening of an additional channel of the current from the Fermi level when an LO-phonon is emitted.

\begin{figure}[ht]
\centering\includegraphics[width=\columnwidth]{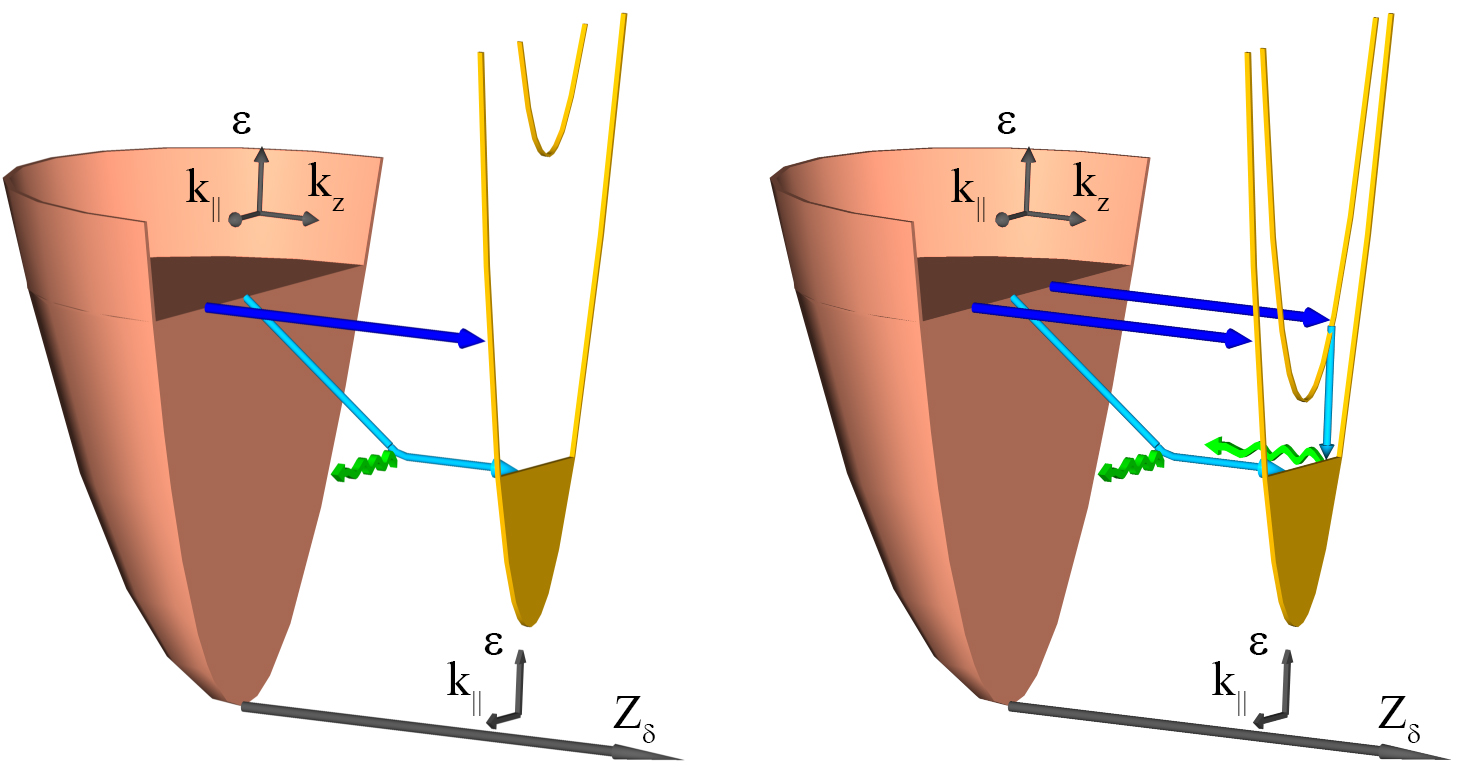}
\caption{
Model of tunneling from a metal electrode (characterized by a 3D spectrum) to a 2D electron system at the threshold of the
emission of an LO-phonon. 
Electrons tunnel from the Fermi surface of the metal to (a) one subband with energy being conserved
(filled arrow) and after the emission of a phonon in the barrier region (open arrow, inelastic tunneling) and (b) to two subbands when,
during tunneling to the upper subband, transitions are possible to the Fermi level of the lower subband with the emission of LOphonons.
This results in back scattering (reflection) of the electrons tunneling from the metal.}
\label{5}
\end{figure}

With the use of external influence, it is possible to move apart the subbands in a $\delta$-layer with the diamagnetic shift in a parallel magnetic field~\cite{4} or to condense them owing to the effect of persistent tunneling photoconductivity (PTPC) discovered in~\cite{17}. 
When, owing to the diamagnetic shift, bottom $E_{1F}$ is pushed up over the threshold to $43$\, meV (Fig.~\ref{4}, closed circles), $\Delta\sigma^{*}$ remains positive. 
Such a situation, when electrons from the Fermi surface of metal tunnel to empty states in the
2DES lower subband, is schematically shown in Fig.~\ref{5}a. 
An additional channel then opens for tunneling within the same subband with the emission of an
LO-phonon in the entire area of the barrier between the metal and the 2DES. 
Therefore, tunnel conductance should increase and the contribution from inelastic electron-phonon scattering within subband $\Delta\sigma_{int}$ should correspond to a positive step on the conductance
curve. 
In this case, $\Delta\sigma_{bg}/\sigma_0$ may be represented as $\Delta\sigma_{int}/\sigma_0$, where 
$\Delta\sigma_{int} > 0$.

After a sample has been exposed to light from an LED during time $t$, energies of empty ($E_i - E_{F\delta} > 0$) levels ''thicken'' to the ground state. 
When the PTPC effect is saturated, the value of $E_{1F}(t)$ in sample 1 falls to $22$\, meV, i.e., to a value that is significantly lower than the threshold (Fig.~\ref{4}, open circles). 
Electrons now tunnel to two subbands of the 2DES. 
As follows from Fig.~\ref{5}b, when second subband $E_1$ ($E_1 - E_0 > \varepsilon_{LO}$) is involved
in the tunnel charge transfer, electrons can, after having emitted an LO-phonon with a nonzero component
of momentum $k_z$ in the direction of tunneling, experience a transition to subband $E_0$. 
As a result, additional contribution to conductance $\Delta\sigma_{ext}$ related to such a
transition of $E_{01}$ proves to be negative. 
This result may be explained in qualitative terms by back scattering (reflection) of tunneling electrons as a result of collision with the LO-phonons emitted during transitions between the subbands in the 2DES. 
In this case, the relative conductance in the vicinity of the threshold is contributed by two mechanisms: $\Delta\sigma_{bg}/\sigma_0 = (\Delta\sigma_{int} + \Delta\sigma_{bg})/\sigma_0$. 
When $|\Delta\sigma_{ext}|$ exceeds $\Delta\sigma_{int}$ , $\Delta\sigma^{*}$ becomes negative, as is observed in the experiment (Fig.~\ref{4}, curve 1). 
Note that $\Delta\sigma^{*}$ crosses the energy axis at $E_{1F} \simeq 33$\, meV, i.e., somewhat below the threshold, since contribution of $\Delta\sigma_{int} > 0$ has to be compensated.

\begin{figure}[ht]
\centering\includegraphics[width=0.8\columnwidth]{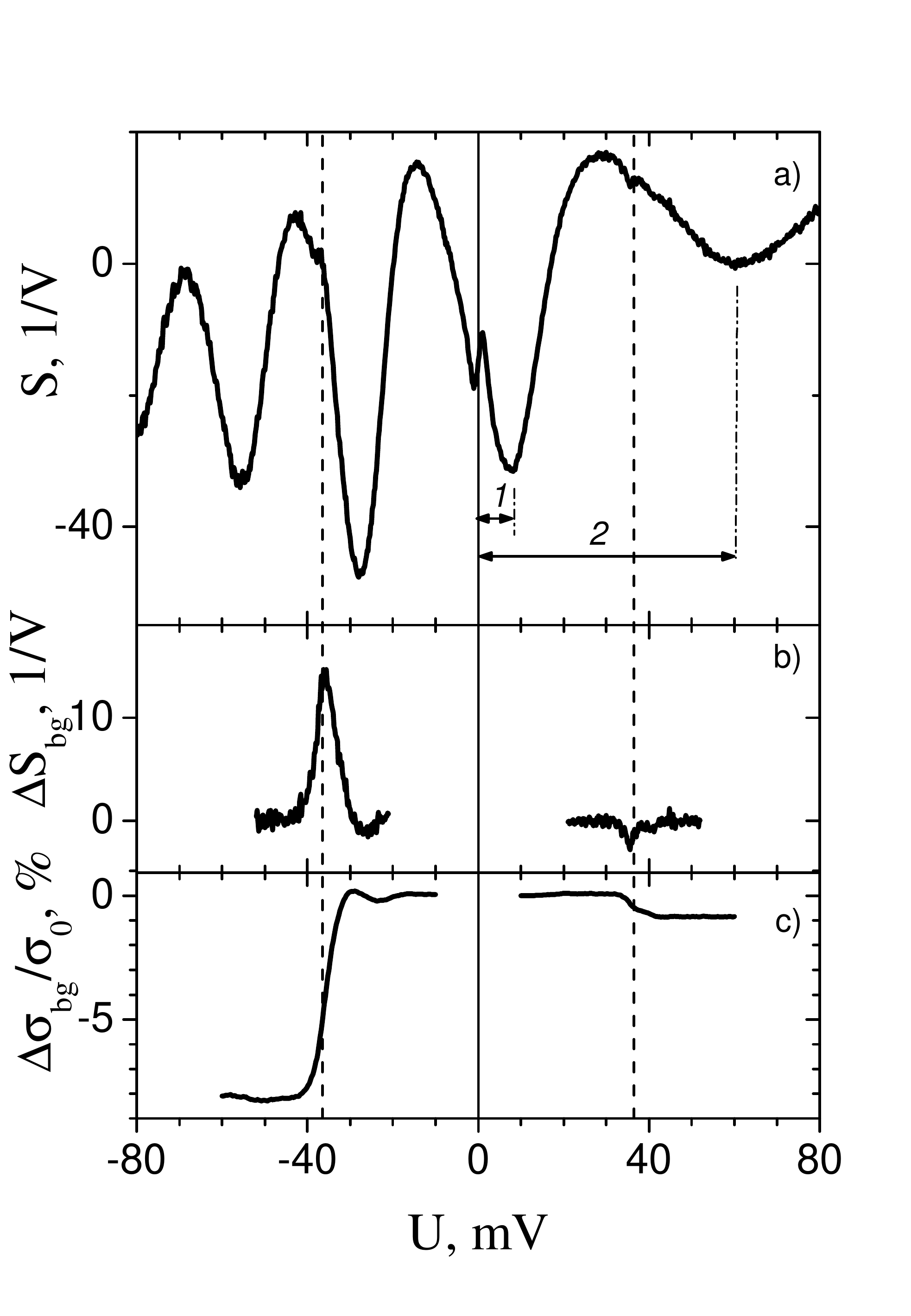}
\caption{
(a) Tunnel spectrum $S$, (b) phonon lines, and (c) relative conductivities in the area of phonon singularities.
Arrows show energies of occupied subbands (1) $E_{1F}$ and (2) $E_{0F}$.
}
\label{6}
\end{figure}

The obtained results show that the emergence of the phonon line with $\Delta\sigma^{*} < 0$, which corresponds to reflection of electrons, is related to a new subband being involved in the tunneling process and to the emission of LO-phonons when electrons experience transitions from this subband to the ground state in the 2DES region. 
This conclusion is in line with calculations of a 3D tunnel system carried out in~\cite{10}, where an analysis of the imaginary part of the electron-phonon system’s self-energy has indicated the existence of the reflection effect during tunneling. 
Additionally, it has been found in~\cite{10} that $\Delta\sigma^{*} < 0$ when relaxation processes are
localized on LO-phonons in one of the tunnel system's electrodes. 
Note that, in sample 1, only usual processes of inelastic tunneling with $\Delta\sigma^{*} > 0$ are observed during tunneling of electrons from the 2DES single occupied subband to a metal electrode. 
This conclusion follows from the shape of the phonon line in the spectrum for this sample, which is shown in Fig.~\ref{2} for $U > 0$. 
The situation changes drastically in the case of sample 2, where reflection effects are observed for both directions of the tunneling process (Fig.~\ref{6}).

\subsection{Reflection of Electrons in the Process of Tunneling to a High-Density 2DES and Near a Polaron Resonance}

As follows from the tunnel spectrum displayed in Fig.~\ref{6}, the energies of the first three levels for sample 2 are $E_{0F} = -61$, $E_{1F} = -7.5$, and $E_{2F} = 28$\, meV. 
Here, subbands $E_1$ and $E_0$ are occupied, a result that corresponds to a concentration of $\approx 2 \times 10^{12}$\, cm$^{-2}$, i.e., to a high-density 2DES with the Fermi energy exceeding $\varepsilon_{LO}$. 
In addition, the phonon lines in this sample differ significantly from those in the case of a low-density 2DES; namely, the values of $\Delta\sigma^{*} < 0$ for both directions of tunneling
(compare the phonon lines in Fig.~\ref{6} with the data for sample 1 shown in Fig.~\ref{2}). 
Let us consider the behavior of the phonon lines in sample 2 starting with the process of tunneling in the 2DES. 
In this case, similar to the case of sample 1, one can assume that the main tunnel current on the threshold of the emission of LO-phonons is related to the tunnel transfer of charges from the metal to the subbands $E_{2F} = 28$\, meV and $E_{1F} = -7.5$\, meV. 
The analysis shows that the probability of tunneling to level $E_0$ is significantly smaller than that
for upper subbands. 
The situation with additional reflection is then implemented in sample 2 and an additional reflection 
($\Delta\sigma^{*} < 0$) (see Fig.~\ref{5}b) is expected to be observed near the threshold. 
Indeed, this is seen in Fig.~\ref{4}, where the closed squares correspond to the initial
tunnel spectrum for $E_{2F} = 28$\, meV and $\Delta\sigma^{*} = -0.08$.
With the exposure of the sample to magnetic field $B_{||}$ parallel to the $\delta$-layer, it is possible to transfer level $E_{2F}$ above the threshold located at $E_{F\delta} + \varepsilon_{LO}$ with the diamagnetic
shift and by moving along the curve shown by closed squares (see Fig.~\ref{4}).

It is seen that $\Delta\sigma^{*}$ vanishes when $E_{2F}(B_{||}) = \varepsilon_{LO}$, i.e.,
when the tunneling process is directed toward occupied subband $E_1$ alone (see Fig.~\ref{5}a). 
The value of $\Delta\sigma^{*}$ then proves to be positive and reaches $\sim 1$\% only when $E_{2F} \simeq
45$\, meV. 
However, in contrast to sample 1, the processes of inelastic tunneling within the subband with
$\Delta\sigma_{int} > 0$ are found to be suppressed strongly. 
This phenomenon is also indicated by the shift of the zero-crossing point to the value $E_{2F}(B_{||}) = \varepsilon_{LO}$, which corresponds to equal contributions of intersubband scattering $|\Delta\sigma_{ext}|$ and of intrasubband scattering $\Delta\sigma_{int}$ by LO-phonons.
Recall that, for sample 1, this point corresponds to $E_{1F}(t) = 33$\, meV $< \varepsilon_{LO}$, where contributions $|\Delta\sigma_{ext}|$ and $|\Delta\sigma_{int}|$ are on the same order of magnitude. 
The specific features of the behavior of the phonon contribution to tunnel conductance discovered in sample 2 require additional study of the reflection processes in the 2DES where density is as high as in this sample. 
(The concentration of 2D electrons is $\approx 2 \times 10^{12}$\, cm$^{-2}$.) 
Here we only highlight the most pronounced features of the interaction between electrons and optic phonons under such conditions.

\begin{figure}[ht]
\centering\includegraphics[width=0.8\columnwidth]{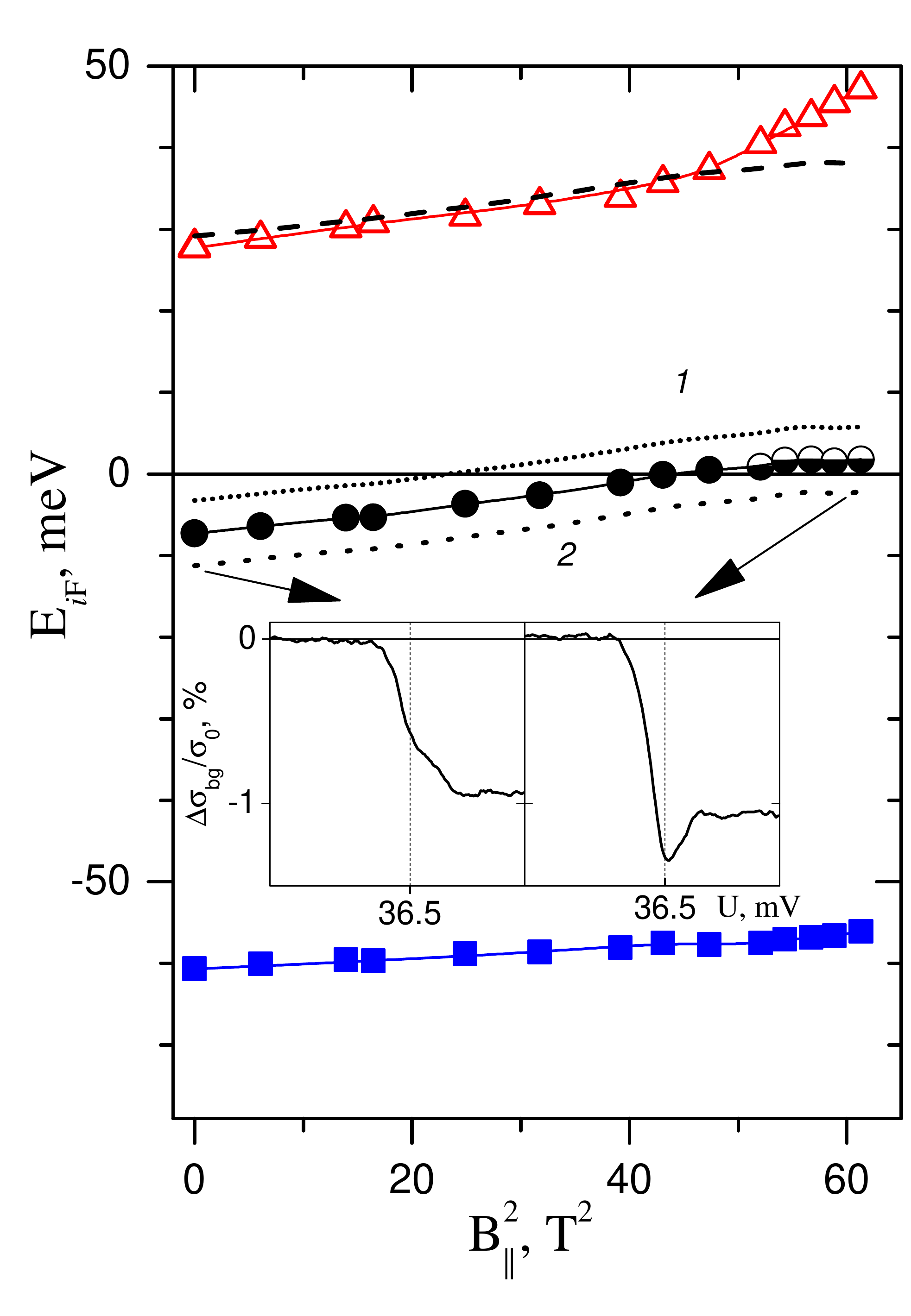}
\caption{
Dependence of subbands' positions in sample 2 on the magnetic field directed along the $\delta$-layer in an Al-$\delta$-GaAs structure. 
Filled squares correspond to $E_{0F}$; circles, to $E_{1F}$; and open triangles, to $E_{2F}$. 
Pinning of subband $E_{2}$ to energy $E_{1F} + \varepsilon_{LO}$ is seen (broken line), thereby indicating
the existence of intersubband polaron interaction. 
The inserts show the dependence of $\Delta\sigma_{bg}/\sigma_0$ on $U$ when tunneling
occurs from the 2DES at the threshold of the emission of an LO-phonon.
}
\label{7}
\end{figure}

Above all, note that the data obtained in a magnetic field for sample 2 show resonance (with respect to the
LO-phonon) interaction between levels $E_1$ and $E_2$. 
The dependence of the positions of the energy levels in the $\delta$-layer on (Fig.~\ref{7}) shows that level $E_2$ follows after $E_1 + \varepsilon_{LO}$ (pinning), thus indicating the existence of the
intersubband polaron interaction~\cite{5}. 
Indeed, as has been observed already for $\delta$-layers~\cite{4}, the diamagnetic shift of the \textit{i}th level is $\Delta E_{i} \sim \Delta Z_{i}$ , where $\Delta Z_i$ is the wave function’s characteristic width in the ith state.
Even though $\Delta Z_{2}/\Delta Z_{1} \simeq 1.5$ according to estimates, the diamagnetic shift of the levels remains unchanged up to $B_{||c} \simeq 7$\, T. 
Thereafter, when grows, variation $\Delta E_2$ increases to the expected values. 
Depletion of subband $E_1$ with the increase of the magnetic field seems to result in a ''collapse'' of the resonance polaron state $E_2 - E_1 = \varepsilon_{LO}$. 
If $B_{||} > B_{||c}$, level $E_1$ is depleted and turns out to be higher than $E_{F\delta}$ (Fig.~\ref{7}, filled circles). 
A selfconsistent calculation (neglecting interaction between electrons and phonons) shows that, for $U = -36.5$\, mV, this phenomenon may occur earlier, even at $B_{||} \simeq 5$\, T (see Fig.~\ref{7}, curve 1). Note that, interestingly, the reflection effects are only observed in the region where level
$E_2$ is pinned to $E_1 + \varepsilon_{LO}$. 
In the region of repulsion of levels (collapse of the resonance polaron state), the contribution of the inelastic-tunneling processes with $\Delta\sigma^{*} > 0$ proves to be negligible as compared to those in
sample 1. 
This may indicate that a relation exists between the intersubband polaron interaction, effects of 
inelastic tunneling, and reflection on the threshold of the emission of an LO-phonon.

The obtained data show that the effects of electron reflection can be observed during tunneling from the
2DES, when electrons experience transitions to the metal from two occupied subbands. 
This is the case for sample 2. 
Curves $\Delta\sigma_{bg}/\sigma_0(U)$ (see Fig.~\ref{7}, insert) show how the conductance decreases for $U > 0$ beyond the threshold. 
Additional experiments are needed to determine the 2DES density for which $\Delta\sigma^{*}$ will change its
sign as a result of tunneling from the $\delta$-layer. 
Remember that $\Delta\sigma^{*} > 0$ is observed reliably for sample 1, where the concentration is $\approx 5 \times 10^{11}$\, cm$^{-2}$. 
Note also that in the experiments presented in Fig.~\ref{7} the maximum magnetic field of $8$\, T proved to be insufficient for pushing up subband $E_1$ from below the 2DES’ Fermi level. 
Although the measured position of $E_{1F}$ is somewhat higher than the Fermi level already at $B_{||c} \simeq 7$\, T, when $U = 36.5$\, mV (the value near which the phonon line is measured), the expected position of $E_{1F}$ should be several millielectron volts lower owing to enrichment of the channel containing the 2DES for the positive bias values on the metal electrode. 
Estimates (a self-consistent calculation with the electron-phonon interaction neglected) show that this downward shift is $\sim 4$\, meV. 
This yields a new position of $E_{1F}$ (Fig.~\ref{7}, curve 2) that turns out to be lower than the Fermi level in the 2DES.

\section{CONCLUSIONS}
\label{s:conc}

Thus, when the concentration of 2D electrons in the near-surface $\delta$-doped layer increases, an Al-$\delta$-GaAs tunnel system exhibits new features of the effects discovered earlier: reflection of electrons at the threshold of the emission of LO-phonons~\cite{13} and intersubband polaron interaction~\cite{4}. 
To study these features further, it is necessary to make special samples whose concentration
of electrons in the $\delta$-layer and population of levels may be varied over a broad range. 
Experiments with such structures are planned for the near future.

\section*{ACKNOWLEDGMENTS}

We thank Yu.V. Fedorov and A.S. Bugaev for the preparation of samples, N.A. Mordovets for help with
conducting measurements, V.A. Kokin for performing self-consistent calculations of the levels in the $\delta$-layer, and S.N. Artemenko, V.A. Volkov, and A.Ya. Shul’man for helpful discussions.

This study was supported by the Russian Foundation for Basic Research (project nos. 03-02-16756-a,
05-02-16316-a, 05-02-17095-a, and 06-02-16955-a) and by the program of basic research of the Division of
Physical Sciences, Russian Academy of Sciences.





\bibliographystyle{elsarticle-num-names}
\bibliography{Reflection}

\begin{thebibliography}{17}
\expandafter\ifx\csname natexlab\endcsname\relax\def\natexlab#1{#1}\fi
\providecommand{\url}[1]{\texttt{#1}}
\providecommand{\href}[2]{#2}
\providecommand{\path}[1]{#1}
\providecommand{\DOIprefix}{doi:}
\providecommand{\ArXivprefix}{arXiv:}
\providecommand{\URLprefix}{URL: }
\providecommand{\Pubmedprefix}{pmid:}
\providecommand{\doi}[1]{\href{http://dx.doi.org/#1}{\path{#1}}}
\providecommand{\Pubmed}[1]{\href{pmid:#1}{\path{#1}}}
\providecommand{\bibinfo}[2]{#2}
\ifx\xfnm\relax \def\xfnm[#1]{\unskip,\space#1}\fi
\bibitem[{Shik(1992)}]{1}
\bibinfo{author}{A.~Y. Shik},
\newblock \bibinfo{title}{Semiconductor structures with $\delta$ layers
  (review)},
\newblock \bibinfo{journal}{Soviet physics. Semiconductors}
  \bibinfo{volume}{26} (\bibinfo{year}{1992}) \bibinfo{pages}{1161}.
\bibitem[{Kotel'Nikov et~al.(1992)Kotel'Nikov, Kokin, Medvedev, Mokerov,
  Rzhanov, and Anokhina}]{2}
\bibinfo{author}{I.~N. Kotel'Nikov}, \bibinfo{author}{V.~A. Kokin},
  \bibinfo{author}{B.~K. Medvedev}, \bibinfo{author}{V.~G. Mokerov},
  \bibinfo{author}{Y.~A. Rzhanov}, \bibinfo{author}{S.~P. Anokhina},
\newblock \bibinfo{title}{Characteristics and special features of the
  conductivity of surface δ-doped layers in gaas with a variable density of
  two-dimensional electrons},
\newblock \bibinfo{journal}{Soviet physics. Semiconductors}
  \bibinfo{volume}{26} (\bibinfo{year}{1992}) \bibinfo{pages}{1462}.
\bibitem[{Schubert(1996)}]{3}
\bibinfo{author}{E.~F. Schubert}, \bibinfo{title}{Delta Doping of
  Semiconductors}, \bibinfo{publisher}{Cambridge Univ. Press, Cambridge},
  \bibinfo{year}{1996}, p. \bibinfo{pages}{498}.
\bibitem[{Kotel'nikov et~al.(2000)Kotel'nikov, Kokin, Fedorov, Hook, and
  Talbaev}]{4}
\bibinfo{author}{I.~N. Kotel'nikov}, \bibinfo{author}{V.~A. Kokin},
  \bibinfo{author}{Y.~V. Fedorov}, \bibinfo{author}{A.~V. Hook},
  \bibinfo{author}{D.~T. Talbaev},
\newblock \bibinfo{title}{Intersubband resonance polarons in al/$\delta$-gaas
  tunneling junctions},
\newblock \bibinfo{journal}{Journal of Experimental and Theoretical Physics
  Letters} \bibinfo{volume}{71} (\bibinfo{year}{2000})
  \bibinfo{pages}{387--390}. \URLprefix \url{https://doi.org/10.1134/1.568361}.
  \DOIprefix\doi{10.1134/1.568361}.
\bibitem[{Levinson and Rashba(1973)}]{5}
\bibinfo{author}{I.~B. Levinson}, \bibinfo{author}{E.~I. Rashba},
\newblock \bibinfo{title}{Bound states of electrons and excitons with optical
  phonons in semiconductors},
\newblock \bibinfo{journal}{Phys. Usp.} \bibinfo{volume}{15}
  (\bibinfo{year}{1973}) \bibinfo{pages}{663--664}. \URLprefix
  \url{https://ufn.ru/en/articles/1973/5/k/}.
  \DOIprefix\doi{10.1070/PU1973v015n05ABEH005043}.
\bibitem[{Liu et~al.(2001)Liu, Cheung, SpringThorpe, Dharma-wardana,
  Wasilewski, Lockwood, and Aers}]{6}
\bibinfo{author}{H.~C. Liu}, \bibinfo{author}{I.~W. Cheung},
  \bibinfo{author}{A.~J. SpringThorpe}, \bibinfo{author}{C.~Dharma-wardana},
  \bibinfo{author}{Z.~R. Wasilewski}, \bibinfo{author}{D.~J. Lockwood},
  \bibinfo{author}{G.~C. Aers},
\newblock \bibinfo{title}{Intersubband raman laser},
\newblock \bibinfo{journal}{Applied Physics Letters} \bibinfo{volume}{78}
  (\bibinfo{year}{2001}) \bibinfo{pages}{3580--3582}. \URLprefix
  \url{https://doi.org/10.1063/1.1377857}. \DOIprefix\doi{10.1063/1.1377857}.
  \href{http://arxiv.org/abs/https://doi.org/10.1063/1.1377857}{{\tt
  arXiv:https://doi.org/10.1063/1.1377857}}.
\bibitem[{Liu et~al.(2003)Liu, Song, Wasilewski, SpringThorpe, Cao,
  Dharma-wardana, Aers, Lockwood, and Gupta}]{7}
\bibinfo{author}{H.~C. Liu}, \bibinfo{author}{C.~Y. Song},
  \bibinfo{author}{Z.~R. Wasilewski}, \bibinfo{author}{A.~J. SpringThorpe},
  \bibinfo{author}{J.~C. Cao}, \bibinfo{author}{C.~Dharma-wardana},
  \bibinfo{author}{G.~C. Aers}, \bibinfo{author}{D.~J. Lockwood},
  \bibinfo{author}{J.~A. Gupta},
\newblock \bibinfo{title}{Coupled electron-phonon modes in optically pumped
  resonant intersubband lasers},
\newblock \bibinfo{journal}{Phys. Rev. Lett.} \bibinfo{volume}{90}
  (\bibinfo{year}{2003}) \bibinfo{pages}{077402}. \URLprefix
  \url{https://link.aps.org/doi/10.1103/PhysRevLett.90.077402}.
  \DOIprefix\doi{10.1103/PhysRevLett.90.077402}.
\bibitem[{Ivanov et~al.(2000)Ivanov, Takhtamirov, Dubrovskii, Volkov, Eaves,
  Main, Henini, Maude, Portal, Maan, and Hill}]{8}
\bibinfo{author}{D.~Y. Ivanov}, \bibinfo{author}{E.~Takhtamirov},
  \bibinfo{author}{Y.~V. Dubrovskii}, \bibinfo{author}{V.~A. Volkov},
  \bibinfo{author}{L.~Eaves}, \bibinfo{author}{P.~C. Main},
  \bibinfo{author}{M.~Henini}, \bibinfo{author}{D.~K. Maude},
  \bibinfo{author}{J.-C. Portal}, \bibinfo{author}{J.~C. Maan},
  \bibinfo{author}{G.~Hill},
\newblock \bibinfo{title}{Observation of the interaction between landau levels
  of different two-dimensional subbands in gaas in a normal magnetic field},
\newblock \bibinfo{journal}{Journal of Experimental and Theoretical Physics
  Letters} \bibinfo{volume}{72} (\bibinfo{year}{2000})
  \bibinfo{pages}{476--479}. \URLprefix
  \url{https://doi.org/10.1134/1.1339904}. \DOIprefix\doi{10.1134/1.1339904}.
\bibitem[{Conley and Mahan(1967)}]{9}
\bibinfo{author}{J.~W. Conley}, \bibinfo{author}{G.~D. Mahan},
\newblock \bibinfo{title}{Tunneling spectroscopy in gaas},
\newblock \bibinfo{journal}{Phys. Rev.} \bibinfo{volume}{161}
  (\bibinfo{year}{1967}) \bibinfo{pages}{681--695}. \URLprefix
  \url{https://link.aps.org/doi/10.1103/PhysRev.161.681}.
  \DOIprefix\doi{10.1103/PhysRev.161.681}.
\bibitem[{Appelbaum and Brinkman(1970)}]{10}
\bibinfo{author}{J.~A. Appelbaum}, \bibinfo{author}{W.~F. Brinkman},
\newblock \bibinfo{title}{Interface effects in normal metal tunneling},
\newblock \bibinfo{journal}{Phys. Rev. B} \bibinfo{volume}{2}
  (\bibinfo{year}{1970}) \bibinfo{pages}{907--915}. \URLprefix
  \url{https://link.aps.org/doi/10.1103/PhysRevB.2.907}.
  \DOIprefix\doi{10.1103/PhysRevB.2.907}.
\bibitem[{11(2003)}]{11}
\bibinfo{title}{Quasi-particle spectrum of 2deg and many-body singularities in
  tunnelling at resonant polaron conditions},
\newblock in: \bibinfo{booktitle}{Proc. 26th Int. Conf., Edinburgh, Scotland,
  UK, 2002}, 171, \bibinfo{publisher}{Institute of Physics Publishing},
  \bibinfo{address}{Edinburgh, Scotland, UK}, \bibinfo{year}{2003}, p.
  \bibinfo{pages}{158}.
\bibitem[{12(2004)}]{12}
\bibinfo{title}{Polaron singularities in tunnelling spectra of high density 2d
  electron system in delta-layer},
\newblock in: \bibinfo{booktitle}{12th Int. Symp. ''Nanostructures: Physics and
  Technology''}, 171, \bibinfo{publisher}{Ioffe Institute},
  \bibinfo{address}{St Petersburg, Russia}, \bibinfo{year}{2004}, pp.
  \bibinfo{pages}{366--367}.
\bibitem[{Kotel'nikov and Dizhur(2005)}]{13}
\bibinfo{author}{I.~N. Kotel'nikov}, \bibinfo{author}{S.~E. Dizhur},
\newblock \bibinfo{title}{Scattering involving lo phonons in tunneling to the
  2d electron system of a delta layer},
\newblock \bibinfo{journal}{Journal of Experimental and Theoretical Physics
  Letters} \bibinfo{volume}{81} (\bibinfo{year}{2005})
  \bibinfo{pages}{458--461}. \URLprefix
  \url{https://doi.org/10.1134/1.1984029}. \DOIprefix\doi{10.1134/1.1984029}.
\bibitem[{Wolf(2012)}]{14}
\bibinfo{author}{E.~L. Wolf}, \bibinfo{title}{Principles of Electron Tunneling
  Spectroscopy}, \bibinfo{publisher}{OUP Oxford}, \bibinfo{year}{2012}.
\bibitem[{Dizhur et~al.(2004)Dizhur, Voronovsky, Fedorov, Kotel'nikov, and
  Dizhur}]{15}
\bibinfo{author}{E.~M. Dizhur}, \bibinfo{author}{A.~N. Voronovsky},
  \bibinfo{author}{A.~V. Fedorov}, \bibinfo{author}{I.~N. Kotel'nikov},
  \bibinfo{author}{S.~E. Dizhur},
\newblock \bibinfo{title}{Transition of the near-surface $\delta$ layer in an
  al/$\delta$(si)-gaas tunnel structure to the insulating state under
  pressure},
\newblock \bibinfo{journal}{Journal of Experimental and Theoretical Physics
  Letters} \bibinfo{volume}{80} (\bibinfo{year}{2004})
  \bibinfo{pages}{433--435}. \URLprefix
  \url{https://doi.org/10.1134/1.1830663}. \DOIprefix\doi{10.1134/1.1830663}.
\bibitem[{Dizhur and Fedorov(2005)}]{16}
\bibinfo{author}{E.~M. Dizhur}, \bibinfo{author}{A.~V. Fedorov},
\newblock \bibinfo{title}{Tunneling dc spectroscopy and digital processing of
  experimental data},
\newblock \bibinfo{journal}{Instruments and Experimental Techniques}
  \bibinfo{volume}{48} (\bibinfo{year}{2005}) \bibinfo{pages}{455--459}.
  \URLprefix \url{https://doi.org/10.1007/s10786-005-0079-x}.
  \DOIprefix\doi{10.1007/s10786-005-0079-x}.
\bibitem[{Dizhur et~al.(2001)Dizhur, Kotel'nikov, Kokin, and Shtrom}]{17}
\bibinfo{author}{S.~Dizhur}, \bibinfo{author}{I.~Kotel'nikov},
  \bibinfo{author}{V.~Kokin}, \bibinfo{author}{F.~Shtrom},
\newblock \bibinfo{title}{2d-subband spectra variations under persistent
  tunnelling photoconductivity conditions in tunnel delta-gaas/al structures},
\newblock \bibinfo{journal}{Physics of Low-Dimensional Structures (PLDS)}
  \bibinfo{volume}{11/12} (\bibinfo{year}{2001}) \bibinfo{pages}{233--244}.
  \URLprefix \url{https://arxiv.org/abs/2008.11471}.

\end{thebibliography}







\end{document}